\documentclass[twocolumn,amsmath,amssymb,prl]{revtex4}
\usepackage{graphicx}% Include figure files
\usepackage{dcolumn}% Align table columns on decimal point
\usepackage{bm}% bold math
\begin{document}
\input{epsf}

\title{The Optimal Cosmic Epoch for Precision Cosmology}

\author{Abraham Loeb}
\affiliation{Astronomy Department, Harvard University, 60 Garden Street, 
Cambridge, MA 02138, USA}

\begin{abstract}

The statistical uncertainty in measuring the primordial density
perturbations on a given comoving scale is dictated by the number of
independent regions of that scale that are accessible to an observer.
This number varies with cosmic time and diminishes per Hubble volume
in the distant past or future of the standard cosmological model.  We
show that the best constraints on the initial power spectrum of linear
density perturbations are accessible (e.g. through 21-cm intensity
mapping) at redshifts $z\sim 10$, and that the ability to constrain
the cosmological initial conditions will deteriorate quickly in our
cosmic future.

\end{abstract}

\pacs{98.80.-k, 98.80.Es, 98.52.-b}
\date{\today}
\maketitle

\paragraph*{Introduction.} The early Universe was characterized by
linear density perturbations with a fractional amplitude
$\vert\delta({\bf r})\vert\ll 1$, thought to have been seeded by
quantum fluctuations during cosmic inflation across a vast range of
scales spanning more than $\sim 26$ orders of magnitude
\cite{Liddle}. As the Universe evolved, these perturbations were
processed by its radiation and matter constituents. Recent surveys
restricted to a small fraction of the total observable volume of the
Universe allowed observers to read off cosmological parameters from
the ``Rosetta stone'' of these density perturbations to an exquisite
precision of a few percent \cite{Mehta,Keisler,WMAP}.

In this paper we consider the fundamental limit to the precision of
cosmological surveys as a function of cosmic time.  Our analysis
provides a global perspective for optimizing future observations, and
for assessing the ultimate detectability limits of weak features such
as non-Gaussianity from inflation \cite{Malda}. Existing cosmological
data sets are far from optimal. For example, the primary anisotropies
of the cosmic microwave background (CMB) \cite{WMAP} sample only a
two-dimensional (last scattering) surface which represents a small
fraction of the three-dimensional information content of the Universe
at the redshift of hydrogen recombination, $z\sim 10^3$.

It is convenient to analyze the density perturbations in Fourier space
with $\delta_{\bf k}=\int d^3r \delta({\bf r}) \exp\{i{\bf k}\cdot{\bf
r}\}$, where $k=2\pi/\lambda$ is the comoving wavenumber. The
fractional uncertainty in the power spectrum of primordial density
perturbations $P(k)\equiv \langle \vert \delta_{\bf k}\vert^2\rangle$
is given by \cite{LW08,Mao08},
\begin{equation}
{\Delta P({\bar k}) \over P({\bar k})} = {1\over \sqrt{N({\bar k})}},
\end{equation}
where the number of independent samples of Fourier modes with
wavenumbers between $k$ and $k+dk$ in a spherical comoving survey
volume $V$ is,
\begin{equation}
dN(k)= (2\pi)^{-2} k^2 V dk,
\end{equation}
with $N({\bar k})$ being the integral of $dN(k)/dk$ over the band of
wavenumbers of interest around ${\bar k}$.

\begin{figure}[th]
\includegraphics[scale=0.4]{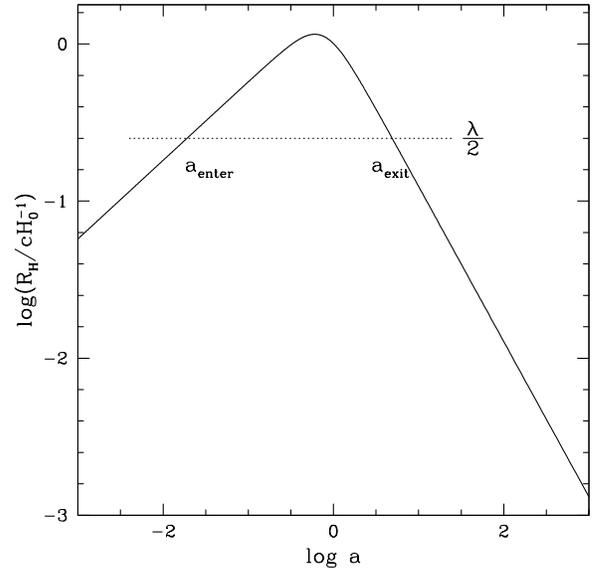}
\caption{In the standard (post-inflation) cosmological model, a
Fourier mode with a comoving wavelength $\lambda$ which enters the
comoving scale of the Hubble radius $R_{\rm H}=c(aH)^{-1}$ (in units
of $cH_0^{-1}=4.3~{\rm Gpc}$) at some early time (corresponding to a
redshift $z=a_{\rm enter}^{-1}-1$), would eventually exit the Hubble
radius at a later time (corresponding to $a_{\rm exit}$). Hence, there
is only a limited period of time when the mode can be observed.}
\label{fig1}
\end{figure}

The maximum comoving wavelength $\lambda_{\rm max}$ that fits within
the Hubble radius is set by the condition (see Figure \ref{fig1}),
\begin{equation}
\lambda_{\rm max}(t)= 2R_{\rm H} ,
\end{equation}
where $R_{\rm H}\equiv c/(aH)$ is the comoving radius of the Hubble
surface, $a=(1+z)^{-1}$ is the scale factor corresponding to a
redshift $z$ (normalized to unity at the present time), and
$H(t)=(\dot{a}/a)$ is the Hubble parameter at a cosmic time $t$.  The
corresponding minimum observable wavenumber $k_{\rm min}(t)\equiv
2\pi/\lambda_{\rm max}$ naturally gives $\int_0^{k_{\rm min}} dk
[dN(k)/dk]\approx 1$. At $z\lesssim 10^3$, $H\approx H_0\sqrt{\Omega_m
a^{-3}+\Omega_\Lambda}$. Throughout the paper, we adopt a present-day
Hubble parameter value of $H_0=70~{\rm km~s^{-1}~Mpc^{-1}}$ (or
equivalently $h=0.7$) and density parameters $\Omega_m=0.3$ in matter
and $\Omega_\Lambda=0.7$ in a cosmological constant \cite{WMAP}.

\begin{figure}[th]
\includegraphics[scale=0.4]{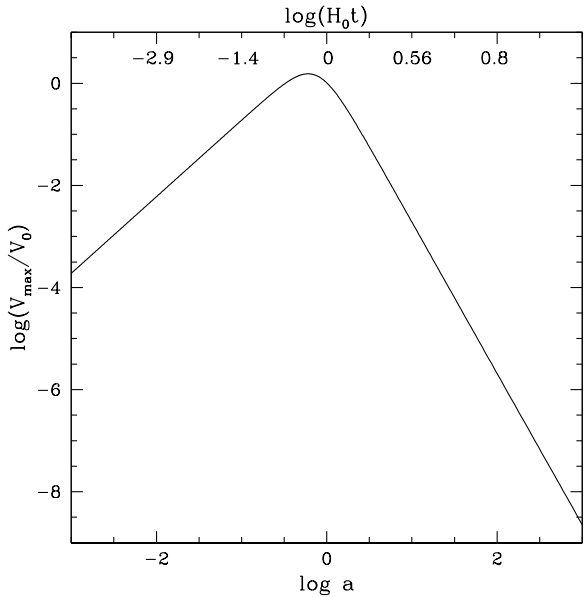}
\caption{The Hubble volume $V_{\rm max}$, normalized by its
present-day value $V_0={4\pi\over 3}(c/H_0)^3=3.3\times 10^2~{\rm
Gpc^3}$, as a function of cosmic time $t$ (top axis, in units of the
Hubble time $H_0^{-1}=14$ Gyr) and scale factor $a=(1+z)^{-1}$ (bottom
axis).}
\label{fig2}
\end{figure}

\paragraph*{Counting Modes.} An observational survey is 
typically limited to a small fraction of the comoving Hubble volume
$V_{\rm max}={4\pi\over 3}R_{\rm H}^3$, which evolves with cosmic
time. Figure \ref{fig2} shows that $V_{\rm max}$ starts small, then
grows to a maximum at the end of the matter dominated era, and finally
diminishes due to the accelerated cosmic expansion.

For the standard cosmological model in which the matter density is
dominated by cold dark matter (LCDM), nonlinear structure develops
first on small spatial scales (large $k$) where it erases memory of
the initial conditions. We associate the maximum wavenumber $k_{\rm
max}$ for which the initial conditions are still in the linear regime
with the minimum radius $R_{\rm NL}=\pi k_{\rm max}^{-1}$ of a
spherical top-hat window for which the root-mean-square amplitude of
density perturbations is unity at the cosmic time of interest
\cite{LF12},
\begin{equation} 
\sigma^2(R_{\rm NL})\equiv \int_0^\infty {4\pi k^2 dk\over (2\pi)^3}
P(k) \left[{3j_1(kR_{\rm NL})\over kR_{\rm NL}}\right]^2 =1 ,
\label{eq:sig}
\end{equation}
where $j_1(x)=(\sin x -x \cos x)/x^2$ is the first spherical Bessel
function. We adopt a present-day normalization of
$\sigma(8h^{-1})=0.8$ for the standard LCDM power-spectrum with a
spectral index $n_s=1$ \cite{WMAP}.

Before nonlinear baryonic structure begins to form ($z\gtrsim 50$) we
associate the corresponding minimum wavelength, $\lambda_{\rm
min}=2\pi/k_{\rm max}$, with the Jeans length for the baryons,
$\lambda_{\rm J}=2\pi/k_{\rm J}$, below which baryonic perturbations
are smoothed out by gas pressure. We base this definition on the
baryons since they are commonly used by observers to trace the
underlying matter distribution. At $z>200$ when the gas is thermally
coupled to the CMB due to the residual fraction of free electrons
after cosmological recombination \cite{LF12},
\begin{equation}
k_J= \left({2k_BT_{\gamma,0}\over 3\mu m_p}\right)^{-1/2}
\sqrt{\Omega_m} H_0 = 0.35~{\rm kpc}^{-1},
\label{kj}
\end{equation}
where $T_{\gamma,0}=2.73$K is the present-day CMB temperature, and
$\mu=1.22$ is the mean atomic weight of neutral primordial gas in
units of the proton mass $m_p$.  At $z\lesssim 200$ and before the
first baryonic objects form, the gas temperature scales as $(1+z)^2$,
and so $k_J$ scales up from the value in Eq. (\ref{kj}) as
$[(1+z)/200]^{-1/2}$. To keep our discussion simple, we ignore
subtleties associated with the motion of the baryons relative to the
dark matter \cite{Tse}, the baryonic temperature fluctuations
\cite{Sma}, and the X-ray and photo-ionization heating of
intergalactic baryons by the first sources of light \cite{PL10}. These
details (which are partially sensitive to the unknown properties of
the first sources of light \cite{LF12}) could reduce $k_{\max}$ at
$z\gtrsim 10$. Hence, our calculated $k_{\rm max}$ should be regarded
as the absolute upper limit, adequate for regions that are not
influenced by these effects.

\begin{figure}[th]
\includegraphics[scale=0.4]{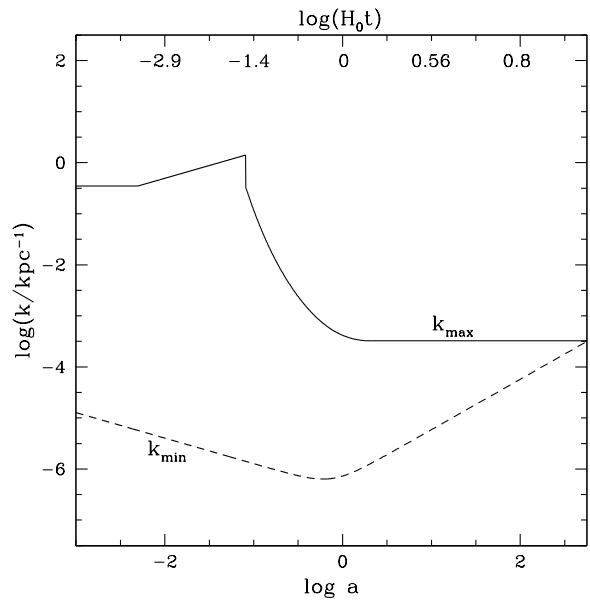}
\caption{The range of comoving wavenumbers for which the linear power
spectrum can be observed per Hubble volume as a function of cosmic
time and scale factor. The minimum wavelength $\lambda_{\rm
min}=2\pi/k_{\rm max}$ is taken as the larger among the baryonic Jeans
scale and the scale where nonlinear structure forms at any given
redshift. The maximum wavelength $\lambda_{\rm max}=2\pi/k_{\rm min}$
is set by the Hubble diameter $2R_{\rm H}$.}
\label{fig3}
\end{figure}

Figure \ref{fig3} shows the range of wavenumbers between $k_{\rm min}$
and $k_{\rm max}$ that are accessible within the Hubble radius as a
function of cosmic time $t$ and scale factor $a(t)$.

The maximum number of linear modes available to an observer at a
cosmic time $t$ is then given by,
\begin{equation}
N_{\rm max} (t)= {V_{\rm max} \over 12\pi^2} \left(k_{\rm max}^3-k_{\rm
min}^3 \right).
\end{equation}

\begin{figure}[th]
\includegraphics[scale=0.4]{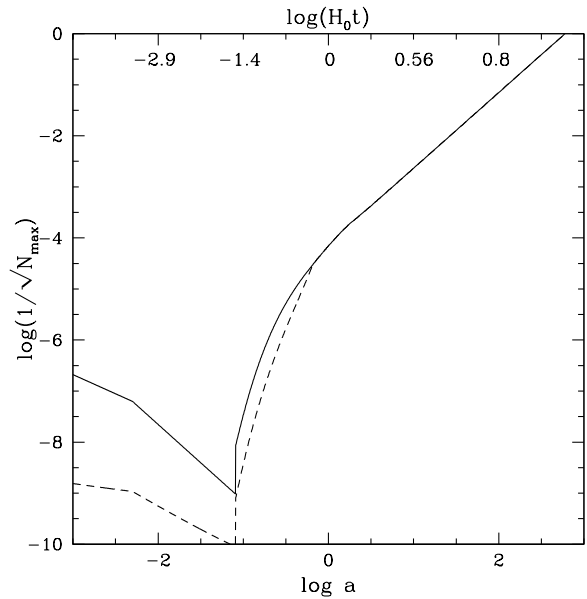}
\caption{The minimum fractional error attainable for the
power-spectrum amplitude $1/\sqrt{N_{\rm max}}$ per Hubble volume, as
a function of cosmic time and scale factor (solid line).  The dashed
line includes the reduction in the statistical uncertainty for a
present-day observer who surveys a spherical shell of comoving width
$2R_H(t)$ centered at the corresponding cosmic times.}
\label{fig4}
\end{figure}

The solid line of Figure \ref{fig4} shows the resulting minimum
fractional error in the power-spectrum amplitude $\Delta P(k_{\rm
max})/P(k_{\rm max})= 1/\sqrt{N_{\rm max}}$, as a function of cosmic
time and scale factor.  We have calculated $R_{\rm NL}$ from
Eq. (\ref{eq:sig}) using the LCDM power spectrum $P(k)$ \cite{WMAP}.
The sharp rise in the minimum error at $a\gtrsim 0.1$ occurs because
the nonlinear scale $R_{\rm NL}$ increases rapidly above the Jeans
length $\lambda_{\rm J}$ around a redshift of $z\sim 10$.  The rise is
sharp since the root-mean-square amplitude of density fluctuations in
LCDM $\sigma(R)$ has a weak $R$-dependence on small scales, implying
that different small scales collapse at about the same cosmic time.
Changing the definition of $R_{\rm NL}$ in Eq. (\ref{eq:sig}) to refer
to $2$--$\sigma$ perturbations reaching an overdensity of unity [i.e.,
$2\sigma(R_{\rm NL})=1$] would simply shift this sharp rise to a value
of $a$ that is smaller by a factor of $2$, since the linear growth
factor of perturbations at $z\gg1$ scales as $a$.

A present-day observer located at a fixed vantage point can see
multiple Hubble volumes of an earlier cosmic time $t$, which were
causally disconnected from each other at that time.  The total number
of such regions available within a spherical shell of comoving width
$\Delta r= 2R_H(t)$ is $N_{\rm regions}=(4\pi r^2 \Delta r)/V_{\rm
max}$, where
\begin{equation}
r(t)=\int_{t}^{t_0} {c dt^\prime\over a(t^\prime)} ,
\end{equation}
is the comoving distance to the shell center for observations
conducted at the present time $t_0$. The reduced statistical
uncertainty of $1/\sqrt{(N_{\rm regions}\times N_{\rm max})}$ is shown
by the dashed line in Figure \ref{fig4}.  Under the most favorable
conditions, the probability distribution of density perturbations can
be compared to a Gaussian form,
$p(\delta)d\delta=(2\pi\sigma^2)^{-1/2}\exp\{-\delta^2/2\sigma^2\}d\delta$,
to within a precision of $\sim 10^{-9}$, several orders of magnitude
better than the level required for detecting non-Gaussianity from
inflation \cite{Malda}. Measurements of the gravitational growth of
$P(k)$ for a particular $k$ at multiple redshifts could test for
modifications of general relativity or hidden constituents of the
Universe to a similar level of precision.

Conventional observational techniques, such as galaxy redshift
surveys, Lyman-$\alpha$ forest spectra, or 21-cm intensity mapping
\cite{LM04,LW08,PL11}, are currently limited by systematic
uncertainties (involving instrumental sensitivity and contaminating
foregrounds) to levels that are well above the ultimate precision
floor presented in Figure \ref{fig4}. But as advances in technology
will break new ground for more precise measurements \cite{exp}, they
might offer a qualitatively new benefit -- enabling observers to
witness the evolution of cosmological quantities [such as $P(k,z)$] in
real time over timescales of years \cite{Sandage62,Loeb98}.

\paragraph*{Conclusions.} Figure \ref{fig4} implies that the most accurate
statistical constraints on the primordial density perturbations are
accessible at $z\sim 10$, when the age of the Universe was a few
percent of its current value (i.e., hundreds of Myr after the Big
Bang). The best tool for tracing the matter distribution at this epoch
involves intensity mapping of the 21-cm line of atomic hydrogen
\cite{LM04,LW08,PL11}.  Although the present time ($a=1$) is still
adequate for retrieving cosmological information with sub-percent
precision, the prospects for precision cosmology will deteriorate
considerably within a few Hubble times into the future. For
simplicity, our quantitative results were derived for a Universe with
a true cosmological constant. However, the ultimate loss of
information holds for any type of accelerated expansion, even if the
dark energy density is evolving in time.

For pedagogical purposes, we considered the instantaneous number of
modes available on a space-like hypersurface of a fixed cosmic time
$t$, under the assumption that a typical cosmological survey would
focus on a small fraction of $V_{\rm max}(t)$ in which the constant
time approximation is valid. Otherwise, the evolution of the density
field and its tracers needs to be taken into account.

Since the past lightcone of any observer covers volumes of the
Universe at earlier cosmic times, one might naively assume that the
accessible information only increases for late-time observers, as past
information will stay recorded near the horizon even in the distant
future \cite{Loeb02,Nagamine,Busha}. However, in practice the
exponential expansion will erase all this information in the future
\cite{Ellis,Island,Loeb02,Chiueh,Gud,Krauss}. Beyond a hundred Hubble
times (a trillion years from now), the wavelength of the CMB and other
extragalactic photons will be stretched by a factor of $\gtrsim
10^{29}$ and exceed the scale of the horizon \cite{HVS} (with each
photon asymptoting towards uniform electric and magnetic fields across
the Hubble radius), making current cosmological sources ultimately
unobservable.  While the amount of information available now from
observations of our cosmological past at $z\sim 10$ is limited by
systematic uncertainties that could potentially be circumvented
through technological advances, the loss of information in our future
is unavoidable as long as cosmic acceleration will persist.

\bigskip
\bigskip
\noindent
\paragraph*{Acknowledgments.}
I thank Charlie Conroy and Adrian Liu for comments on the
manuscript. This work was supported in part by NSF grant AST-0907890
and NASA grants NNX08AL43G and NNA09DB30A.

\end{document}